\documentclass[usenatbib,usegraphicx,useAMS]{mn2e}
\usepackage{amssymb,url,times}

\def\be{\begin{equation}}
\def\ee{\end{equation}}
\def\no{\noindent}

\title[Tidal disruption events]{Multiband lightcurves of tidal disruption events}

\author[Lodato \& Rossi]{Giuseppe Lodato$^{1}$\footnote{giuseppe.lodato@unimi.it}, Elena M. Rossi$^{2}$ \\
$^1$Dipartimento di Fisica, Universit\`a Degli Studi di Milano, Via Celoria, 16, Milano, 20133, Italy\\
$^2$  Racah Institute of Physics, The Hebrew University, Jerusalem 91904, Israel 
}
\date{Submitted: Revised:  Accepted:}
\pagerange{\pageref{firstpage}--\pageref{lastpage}} \pubyear{2010}

\begin{document}
\label{firstpage}
\bibliographystyle{mn2e}
\maketitle

\begin{abstract}
Unambiguous detection of  the tidal disruption of a star would allow an
assessment of the presence and masses of supermassive black holes in
quiescent galaxies. It would also provide invaluable information on
bulge scale stellar processes (such as two-body relaxation) via the
rate at which stars are injected into the tidal sphere of influence of
the black holes. This rate, in turn, is essential to predict
gravitational radiation emission by compact object inspirals. The {\em
  signature} of a tidal disruption event is thought to be a fallback
rate for the stellar debris onto the black hole that decreases as
$t^{-5/3}$. This mass flux is often assumed to yield a luminous signal
that decreases in time at the same rate. In this paper, we calculate
the monochromatic lightcurves arising from such an accretion event.
Differently from previous studies, we adopt a more realistic description 
of the fallback rate and of the super-Eddigton accretion physics.  We 
also provide simultaneous lightcurves in optical, UV and
X-rays.  We show that, after a few months, optical and UV lightcurves
scale as $t^{-5/12}$, and are thus substantially flatter than the $t^{-5/3}$
behaviour, which is a prerogative of the bolometric lightcurve,
only. At earlier times and for black hole masses $< 10^7~M_{\sun}$, the
wind emission dominates: after reaching a peak of $10^{41}-10^{43}$
erg/s at roughly a month, the lightcurve decreases steeply as $\sim
t^{-2.6}$, until the disc contribution takes over. The X-ray band,
instead, is the best place to detect the $t^{-5/3}$ ``smoking gun''
behaviour, although it is displayed only for roughly a year, before the
emission steepens exponentially.
\end{abstract}


\section{Introduction}
\label{sec:intro}

Luminous flares from quiescent (non-AGN) galaxies are often interpreted as
arising from the tidal disruption of stars as they get close to a dormant
supermassive black hole in the centre of the galaxy 
\citep{komossa99,gezari08,gezari09,esquej08,cappelluti09}. 
 If this interpretation is correct, such events represent a very important probe for black holes in quiescent galaxies. 
In principle, detailed modeling of individual events could yield constraints on black hole properties.
More ambitiously, a measurement of the rate of events in galaxies of different types would provide information on the stellar dynamics at parsec scales, that is responsible for injecting stars into orbits that result in tidal disruption  (e.g., \citealt{SU99,MT99}). Knowledge of these stellar dynamical processes is of broad interest, since similar processes contribute to the rate of at which compact objects are captured by (and subsequently merge with) the central black hole \citep{SR97}. These extreme mass ratio inspirals are important targets for future gravitational wave experiments.
Similar processes also occur for lower mass objects, such as in compact binary
systems, hosting i) a stellar mass black hole and a white dwarf \citep{rosswog08}; or ii) two white dwarfs \citep{rosswog07,RB09}.

The pioneering works by \citet{lacy82,rees88,phinney89b} and, from a numerical
point of view, by \citet{evans89} have set the theoretical standard for the
interpretation of such events. In particular, a distinctive feature of this
theory is that, to a first approximation, the rate at which the black hole is fed with the stellar debris after the tidal disruption (the ``fallback rate'' $\dot{M}_{\rm fb}$) should decrease as\footnote{Note that the original paper by \citealt{rees88} quotes a $t^{-5/2}$ dependence, later corrected to $t^{-5/3}$ by \citealt{phinney89b}.}
 $\dot{M}_{\rm fb}(t)\propto t^{-5/3}$. {\it If} the fallback rate can be directly translated into an accretion luminosity, one could then expect the light curve of such events to follow the same behaviour with time. Indeed, a $t^{-5/3}$ light curve is
generally fitted to the observed luminosities of events interpreted as stellar
disruptions \citep{gezari08,gezari09,esquej08,cappelluti09}. In practice, the $t^{-5/3}$ evolution of the fallback rate is only valid several months after the event, and the initial evolution generally shows a relatively gentle rise with time, dependent on the internal structure of the disrupted star \citep{LKP09}. 

However, it is not obvious {\it a priori} that one can make such a direct translation of fallback rate into observed luminosity, for a number of reasons. Firstly, for small black hole masses and at optical frequencies \citet{strubbe09} showed that the emission can contain at early times a significant component arising from a radiatively driven wind, consequence of an inital super-Eddington fallback rate. Additionally, the unbound stellar debris might also give some contribution at optical wavelengths \citep{kasen10}. Secondly,  observations are typically made in some narrow bands. Since the system is intrinsically time variable, this implies that the monochromatic lightcurves should not necessarily scale at all times with the bolometric curve. 

In this paper, we consider systematically both effects and we show that in general the expected monochromatic lightcurve in most bands does not scale as $t^{-5/3}$. When the luminosity is dominated by the accretion of the fallback material, at optical and UV frequencies the lightcurve is proportional to $t^{-5/12}$ during an extended period of time. Only in X-rays one should expect a lightcurve close to the naive expectation and only for a brief period of time. Additionally, during the first year or so since disruption, the wind emission may further complicate the shape of the lightcurve in the optical and in the UV bands. We show that the strength of this contribution depends on the internal structure of the disrupted star and on the fraction of mass that the black hole can accrete, when fed at a super-Eddington rate.

The paper is organized as follows. In section \ref{sec:basic}, we review the basic dynamics involved in a tidal disruption event. In section \ref{sec:scalings}, we demonstrate analytically what are the expected monochromatic lightcurves of the disc and of the wind emission, separately. In section \ref{sec:lightcurve}, we discuss the full lightcurve in several bands, as a function of the main parameters of our model. In section \ref{sec:discussion} we discuss our results and draw our conclusions. 

\section{Basic dynamics}
\label{sec:basic}

Let us consider a star with mass $M_{*}= m_* M_{\odot}$ and radius
 $R_* = x_{\star} R_{\odot}$ which moves in a highly eccentric orbit under
 the sole influence of a supermassive black hole with mass $M = 10^6 M_6
 M_{\odot}$.  If the distance of closest approach $r_{\rm p}$ to the
 black hole is within the tidal sphere of radius
\be
r_{\rm t} = \left(\frac{M}{M_*}\right)^{1/3}R_* \approx 0.47~{\rm AU}~  \left(\frac{M_6}{m_*}\right)^{1/3}x_{\star},
\label{eq:rt}
\ee  
then the star self-gravity is not able to counteract the tidal pull and the star is torn apart.
We define the penetration factor as the ratio of the two distances:
\be
\beta \equiv \frac{r_{\rm t}}{r_{\rm p}}.
\ee
When needed, our standard choice of parameters for the stellar properties is
$\beta=x_{\star}=m_*=1$.

For simplicity, we will assume the star trajectory to be
parabolic. However, the dynamics described here is more generally
valid for stars whose orbital energy is much smaller than the energy
spread caused by tidal forces. 
The stars that enter the tidal sphere
are generally expected to satisfy this requirement, since they should have very
eccentric orbits.

At distances $r$ larger than $r_{\rm t}$, $R_*/r \leq
(M_*/M)^{1/3} \approx 10^{-2}$ and the star behaves effectively as a
point mass in the black hole gravitational field.  However, when the star
approaches its pericenter $r_{\rm p}\approx r_{\rm t}$, the star size becomes
important. The different distances to the black hole at which different fluid
elements lie cause a sizeable spread $\Delta \epsilon$ in specific
energy $\epsilon$ within the star,
\be
\Delta \epsilon \simeq \pm \frac{GM}{r_{\rm p}^2} R_*,
\label{eq:de}  
\ee
\citep{LKP09}, which can greatly exceed the star binding energy $\Delta \epsilon/(G M_*/R_*) =\beta^2 (M/M_*)^{1/3}$.
As a result of this spreading, half of the star debris remains bound
($\epsilon <0$) to the black hole, while the rest ($\epsilon >0$) is
lost from the system.\footnote{If the stellar orbit is not exactly parabolic, then the ratio of bound versus unbound material may vary.} The material which remains bound is ejected in highly eccentric orbits and after a time
\be
\begin{array}{ll}
   t_{\rm min} &=\displaystyle\frac{ 2\pi  G M}{(-2\, \Delta \epsilon)^{3/2}} \\
                           &= \displaystyle\frac{\pi}{2^{1/2}} \left(\frac{r_{\rm p}} {R_*}\right)^{3/2} \sqrt{\frac{r_{\rm p}^3}{G M}}=\\
                           & \approx 41\;~ M_{6}^{1/2} m_*^{-1} \beta^{-3}\,x_{\star}^{3/2}~{\rm d},   \\
   \end{array}
\label{eq:tmin}
\ee
it starts coming back to pericenter at a rate $\dot{M}_{\rm fb}$, which depends on the distribution of specific energies.
If the distribution in energy $\mbox{d}M_*/\mbox{d}\epsilon$ is flat, then we find the classic result \citep{rees88,phinney89b}
\be
\dot{M}_{\rm fb} = \dot{M}_{\rm p} \left(\frac{t}{t_{\rm min}}\right)^{-5/3}.
\label{eq:mdot}
\ee
\no
The peak fallback rate 
\be
\dot{M}_{\rm p} = \frac{1}{3} \frac{M_*}{t_{\rm min}} \approx 1.9 ~ 10^{26} ~M_6^{-1/2} m_*^2~\beta^3 x_{\star}^{-3/2}~{\rm g/s},
\label{eq:mdot2}
\ee
 can be compared to the Eddington accretion rate
\be
\dot{M}_{\rm Edd} = 1.3 ~ 10^{24} M_6~(\eta/0.1)^{-1}$ g s$^{-1},
\ee
where $\eta\approx 0.1$ is the radiative efficiency of the accretion process. We thus see that for a $10^6M_{\odot}$ black hole, the fallback rate can be as large as 100 times the Eddington rate, and is therefore expected to produce a luminous flare, which can be in principle detectable. Note that the ratio $\dot{M}_{\rm p}/\dot{M}_{\rm \rm Edd}\propto M^{-3/2}$, so that the fallback rate is expected to be only marginally super-Eddington for a $10^7M_{\odot}$ black hole. Although also in this case a luminous flare is expected to be produced, its appearence will be significantly different, as discussed below in Section 4.  After a time
\be
t_{\rm Edd} = 760 ~(\eta/0.1)^{3/5} M_6^{-2/5} m_*^{1/5} \beta^{-6/5} x_{\star}^{3/5} \mbox{d},
\ee
the fallback rate ceases to be super-Eddington. 

This simple picture has been extensively tested numerically \citep{NK82,evans89,laguna93b,ayal2000}. In particular, \citet{LKP09} show two important results, that we will use in the following.
First, even when $\beta=1$ and the the spin up of the star is
expected to approach the break up velocity, eq.~(\ref{eq:de}) is a
good estimate for the total energy spread.  Second, the energy
distribution $\mbox{d}M_*/\mbox{d}\epsilon$ within the bound matter
depends on the internal structure of the star (parameterized by its
polytropic index $\gamma$).  In particular they find that
$\mbox{d}M_*/\mbox{d}\epsilon$ is constant only for an idealised
incompressible star ($\gamma \gg 1$). In this idealised case
$\dot{M}_{\rm fb}$ follows eq.~(\ref{eq:mdot}) at all times. In
general, however, the resulting $\dot{M}_{\rm fb}$ departs from the
$t^{-5/3}$ power-law behaviour, during the first few months (see their
figure 10, left panel), resulting in a smaller peak
fallback rate. The departure is more pronounced for more compressible
stellar models.

As matter falls back to $r_{\rm p}$, we assume that it undergoes
 shocks that convert all its kinetic energy into internal
 energy. Initially, when the accretion rate is highly super Eddington,
 only a fraction ($1-f_{\rm out}$) of the fallback material forms a
 disc and can accrete all the way down to the black hole. The
 remaining part of the stellar debris will instead leave the system
 under a strong radiative pressure.  It is not possible to quantify
 analytically the parameter $f_{\rm out}$, because the amount of mass
 that is ejected will depend upon the detailed redistribution of
 energy and angular momentum within the flow \citep{bb04}.  Rather,
 simulations are required to assess the rate at which a hole can be
 fed in such extreme conditions.
\citet{strubbe09}  assume a constant $f_{\rm out}=0.1$. 
For moderate super-Eddington accretion rates, however, numerical slim disc models by \citet{dotan10} indicate that $f_{\rm out}$ is a growing function of $\dot{M}_{\rm fb}/\dot{M}_{\rm Edd}$, reaching $f_{\rm out}\approx 0.7$ for $\dot{M}_{\rm fb}/\dot{M}_{\rm Edd}=20$. For higher accretion rates, they could not find steady state solutions. This does not mean that solutions with a different accretion geometry do not exist. However, it does mean that any extrapolation of their results to more vigorous accretion regimes, that may be attained as a consequence of a tidal disruption event,
should be considered as indicative. Mindful of these uncertainties, we will explore the consequences of different prescriptions for $f_{\rm out}$.

 The material which is not ejected in the wind accretes onto the black hole through a relatively thick disc. Initially, the disc viscous time is much shorter than the timescale at which the material falls back to $r_{\rm p}$. In this phase the rate at which the black hole accretes mass is thus the same as the rate at which the disc is fed from the fallback material. At very late times (at $t\gtrsim 100$ years after the event, e.g. \citealt{ulmer99}) the viscous time becomes long enough that the accretion rate is controlled by viscous processes rather than by the disc feeding. \citet{cannizzo90} have discussed these late stages and found that the accretion rate onto the hole scales as $t^{-1.2}$. In this paper we will not consider these late stages, focussing instead on the early evolution of the system.

\section{Scalings for the multiwavelength emission}
\label{sec:scalings}
In this section, we discuss the relevant dynamics and radiative
properties of the disc and the wind separately.  In particular, we
will focus on the temporal behaviour of monochromatic lightcurves,
which are derived analytically assuming a pure $t^{-5/3}$ power-law for
the accretion rate. Therefore these scalings should be regarded as
being only suggestive at early times ($< $ half a year). However, they
certainly provide a good description at times larger than one year
after disruption, when the observed emission is dominated by the
disc at all frequencies (see Section 4).

\begin{figure*}
\begin{center}
\includegraphics[width=0.65\columnwidth]{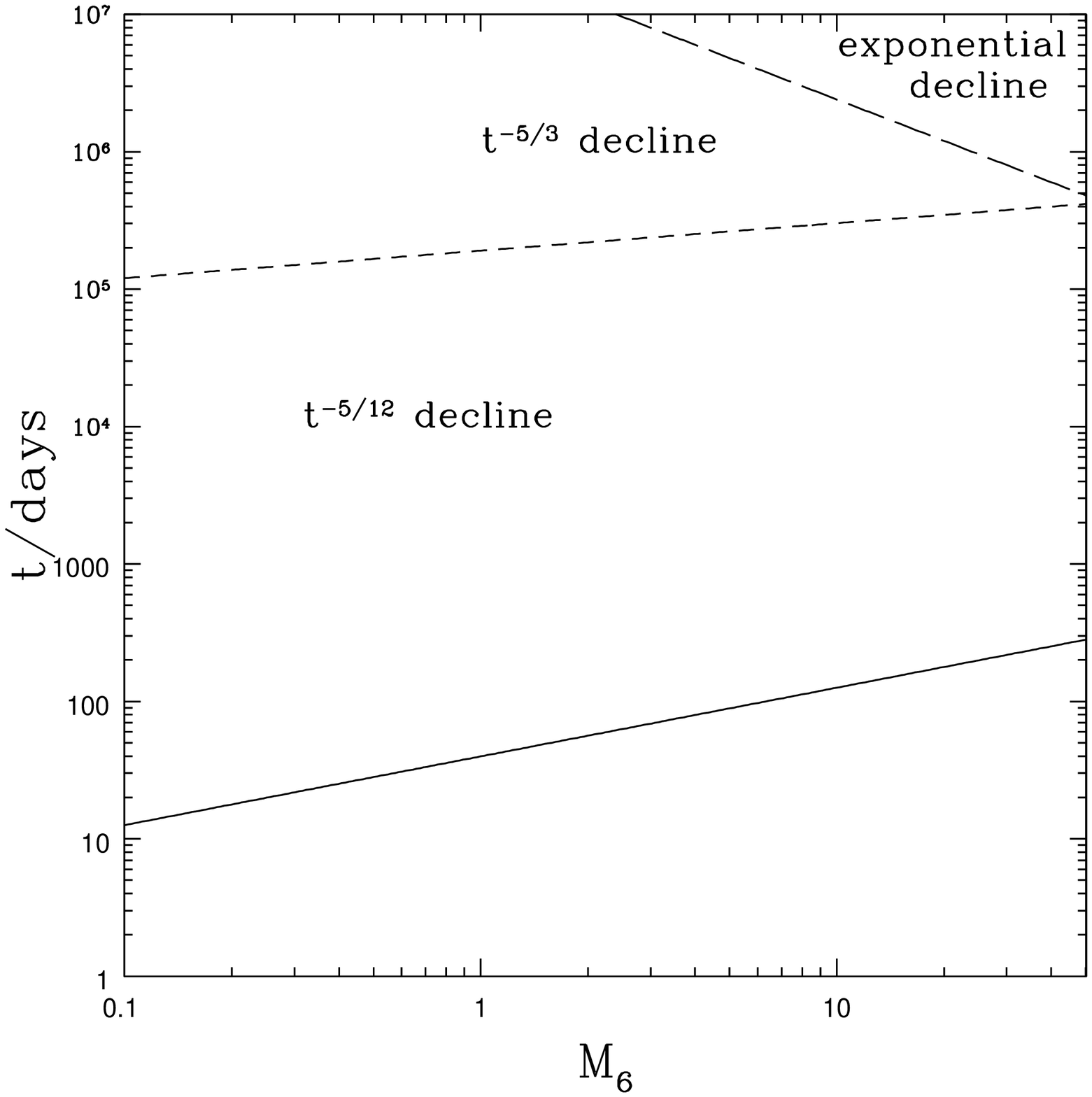}
\includegraphics[width=0.65\columnwidth]{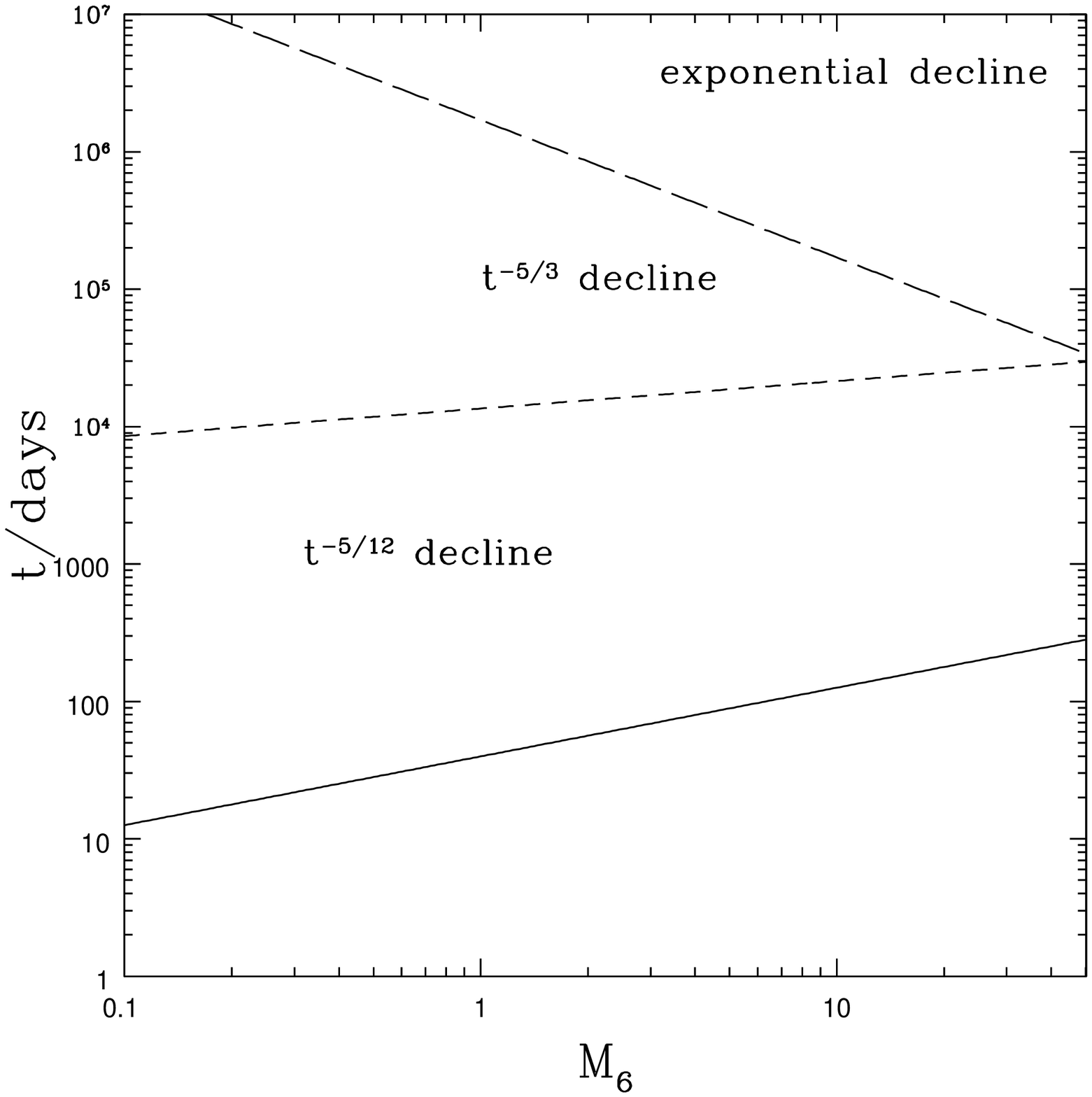}
\includegraphics[width=0.65\columnwidth]{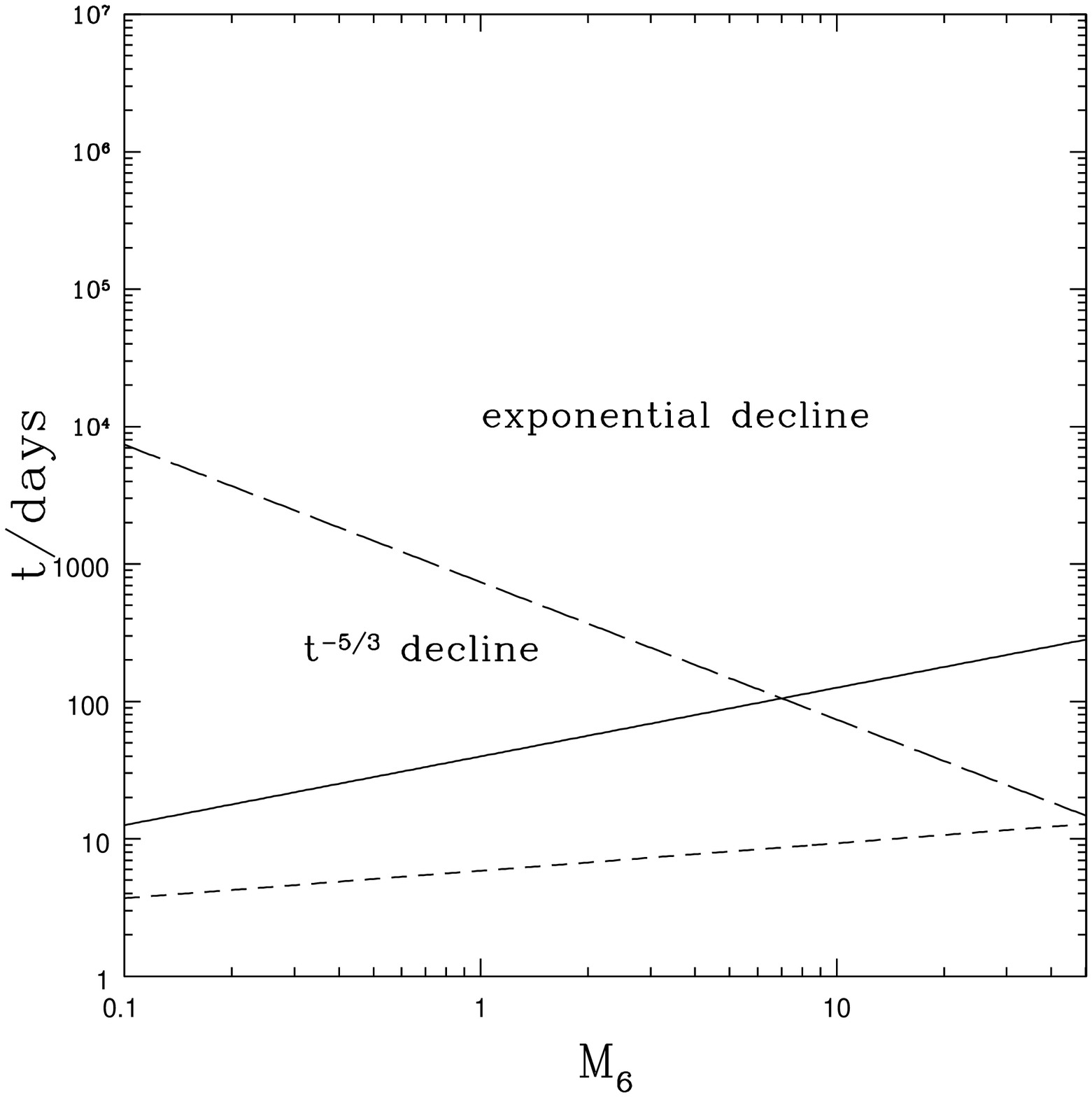}
\caption{Typical timescales as a function of black hole mass. The
  solid line shows $t_{\rm min}$ (Eq. \ref{eq:tmin}), the short dashed line shows $t_{\rm
    ab}$ and the long dashed line shows $t_{\rm bc}$. The three panels
  refer to an observed frequency in the optical g-band (left), in the
  GALEX FUV band (center) and in the X-rays, at 0.2 keV (right). In
  the optical and FUV, for an extended period of time after the event
  the lightcurve is expected to scale as $t^{-5/12}$ before steepening
  into $t^{-5/3}$. It is only in X-rays that the initial lightcurve
  follows the standard $t^{-5/3}$ law.}
\label{fig:times}
\end{center}
\end{figure*}

\subsection{Disc contribution}

In this section, we focus on the disc emission, which is expected to
dominate at high frequencies and/or at late times.  For simplicity, we
adopt here a thin disc geometry, and defer to section
\ref{sec:lightcurve} a discussion of a probably more realistic slim
disc model. For a geometrically thin and optically thick disc, the
effective temperature at a given radius $R$ is related to the mass
accretion rate through:
\begin{equation}
\sigma_{\rm SB}T_{\rm d}^4 = \frac{3GM\dot{M}}{8\pi R^3}\left(1-\sqrt{\frac{R_{\rm in}}{R}}\right),
\end{equation}
where $M$ is the black hole mass, $\dot{M}=(1-f_{\rm out})\dot{M}_{\rm
  fb}$ is the accretion rate, $R_{\rm in}\approx 3R_{\rm S}$ (where
$R_{\rm s}=2GM/c^2$ is the Schwarzshild radius of the black
hole) and $\sigma_{\rm SB}$ is the Stefan-Boltzmann constant.  As time
increases, the accretion rate decreases as $t^{-5/3}$. Thus, while the bolometric 
luminosity scales as $\dot{M}$, the disc temperature decreases as $\dot{M}^{1/4}$.

For a star of mass $M_{\star}=M_{\odot}$ and radius
$R_{\star}=R_{\odot}$, disrupted by a $10^{6}M_{\odot}$ black hole,
the tidal radius is
approximately at $20R_{\rm S}$ from the hole (eq.~\ref{eq:rt}). Since the disc outer
radius is expected to be close to the circularization radius $R_{\rm
  out}=2r_{\rm p} =2r_{\rm t}/\beta$, the radial extent of the disc is
rather small. In general, we have
\begin{equation}
\frac{R_{\rm out}}{R_{\rm in}} \simeq 15~\beta^{-1}M_6^{-2/3}m_{\star}^{-1/3}x_{\star}.
\end{equation}

The spectrum of an optically thick disc is approximately given by a
superposition of blackbody spectra with temperature ranging from
$T_{\rm out}=T(R_{\rm out})$ to $T_{\rm in}=T(R_{\rm in})$. Since the
radial extent of the disc is small, the range of disc temperatures is
correspondingly small: 
\be 
\frac{T_{\rm in}}{T_{\rm out}} \simeq
\left(\frac{R_{\rm out}}{R_{\rm in}}\right)^{3/4} \approx 7.6~
\beta^{-3/4}M_6^{-1/2}m_{\star}^{-1/4}x_{\star}^{3/4}.  
\ee

If the event is observed at a given wavelength $\nu_{\rm obs}$, we may
define a corresponding temperature through Wien's displacement law:
\begin{equation}
T_{\rm obs}\approx 1.7~10^4 \nu_{\rm 15} K,
\label{eq:tobs}
\end{equation}
where we adopt the notation: $\nu_{\rm x}=\nu_{\rm
  obs}/10^{x}$~Hz. The shape of the lightcurve is determined by the
relative size of $T_{\rm obs}$, $T_{\rm in}$ and $T_{\rm out}$. We can
identify three phases:
\begin{itemize}
\item[(a)] $T_{\rm obs}<T_{\rm out}$. This corresponds to the earliest
  times (when the disc is hotter), for relatively small observed
  frequencies. In this case the observed frequency sits in the
  Rayleigh-Jeans tail of the disc spectrum and thus the luminosity
  scales with time in the following way:
\begin{equation}
\nu L_{\nu} \propto T_{\rm d} \propto \dot{M}^{1/4} \propto t^{-5/12}.
\end{equation}
\item[(b)] $T_{\rm out} < T_{\rm obs} < T_{\rm in}$. In this phase the
  emission at $\nu_{\rm obs}$ is dominated by the disc radii for which
  $T(R)\approx T_{\rm obs}$, whose flux is close to the bolometric
  flux. Since the emitting area does not vary much with time 
  (at most by a factor $R_{\rm out}/R_{\rm in}\approx 10$), it is in
  this phase that we can expect a monochromatic luminosity
  proportional to $t^{-5/3}$.
\item[(c)] $T_{\rm in} < T_{\rm obs}$. This phase occurs at late
  times, when the disc has cooled significantly. In this case, the
  observed frequency sits in the Wien part of the spectrum and we
  should expect an exponentially declining monochromatic lightcurve.
\end{itemize}
The transition between the different phases occurs at times $t_{\rm
  ab}$ and $t_{\rm bc}$, such that $T_{\rm out}(t_{\rm ab}) = T_{\rm
  obs}$ and $T_{\rm in}(t_{\rm bc}) = T_{\rm obs}$. These two
transition times are given by: 
\be 
t_{\rm ab} \approx 6.3~10^4 \beta^{3/5}
M_6^{1/5}m_{\star}^{4/5}x_{\star}^{-6/5}\nu_{15}^{-12/5} \mbox{d} ,
\label{eq:tab}
\ee 
\be 
t_{\rm bc} \approx 8.4~10^6\beta^{-6/5}
M_6^{-1}m_{\star}^{1/5}x_{\star}^{3/5}\nu_{15}^{-12/5} \mbox{d} .
\label{eq:tbc}
\ee 
Typically these transitions occur at times much larger than $t_{\rm Edd}$, and we have thus assumed $f_{\rm out}=0$ in Eqs. (\ref{eq:tab}) and (\ref{eq:tbc}). We plot in Figure
\ref{fig:times} the relevant timescales as a function of the black hole mass $M_6$ for three choices of the observed frequency $\nu_{\rm
  obs} = 6.3~10^{14}$ Hz  (optical g-band),
$\nu_{\rm obs} = 1.9~10^{15}$ Hz (the GALEX FUV band)
and $\nu_{\rm obs} = 4.8~10^{16}$ Hz (soft X-rays at $0.2$ keV). The
solid line shows the time $t_{\rm min}$, which marks the beginning of
the tidal disruption event. The short dashed line indicates $t_{\rm
  ab}$ and the long dashed line indicates $t_{\rm bc}$. We thus
clearly see that in the optical and in the FUV, the lightcurve
initially follows a $t^{-5/12}$ lightcurve for an extended period of
time, lasting several hundred years for the g-band and several decades
for the GALEX FUV band. Only in the X-rays is the initial lightcurve
expected to show the standard $t^{-5/3}$ decline for a few years,
before turning into an exponential decline. Note also that the time
span during which we expect a monochromatic $t^{-5/3}$ decline gets
smaller for higher black hole masses, at all frequencies.

\begin{figure}
\begin{center}
\includegraphics[width=\columnwidth]{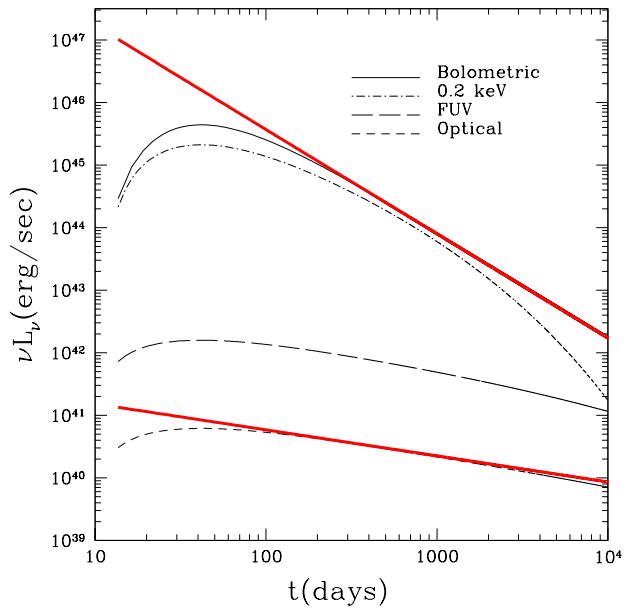}
\caption{Lightcurves for the disc emission from the disruption of a solar type star by a $10^6M_{\odot}$ black hole at 
$\beta=1$. The fallback rate corresponds to a $\gamma=1.4$ polytropic model for the star \citep{LKP09}. The time $t=0$ corresponds to the pericenter passage of the disrupted star. 
The solid line shows the bolometric luminosity, the dot-dashed line shows the luminosity at 0.2 keV, the long-dashed line shows the luminosity in the FUV, at $\nu = 1.9~10^{15}$ Hz, while the short-dashed line shows the luminosity in the optical at $\nu = 6.3~10^{14}$ Hz. The two red lines mark the simple power laws expected for the bolometric luminosity ($\propto t^{-5/3}$) and for the monochromatic luminosity in the optical/UV ($\propto t^{-5/12}$).}
\label{fig:discbolo}
\end{center}
\end{figure}

An example of a disc lightcurve at different frequencies is shown in
Fig. \ref{fig:discbolo}, which corresponds to a solar type star
disrupted by a $10^6M_{\odot}$ black hole.
The fallback rate is the one predicted by \citet{LKP09} for
 a $\gamma=1.4$ polytropic model for the star. 
The two red lines trace, for comparison, a $t^{-5/3}$ and a $t^{-5/12}$
decline. This figure confirms the simple estimates described above:
the optical and FUV lightcurve show initially a $t^{-5/12}$ decline,
while the X-ray lightcurve follows the bolometric $t^{-5/3}$
curve; after a few years the X-ray lightcurve steepens into an
exponential decline.

\subsection{Wind contribution}
\label{sec:wind}

In this section, we derive the monochromatic luminosity scalings for the
wind contribution to the emission that may arise from a tidally disrupted  star.

The main properties of a radiation-driven wind powered by fallback matter
have been calculated by Rossi \& Begelman (2009) and applied to the problem at hand 
by \citet{strubbe09}. Here we state some of their main results
without derivation and we refer the reader to that literature for more details.
In addition, we derive scalings to better illustrate the behaviour of a monochromatic lightcurve.

The main assumptions are the following.
From the launching radius $r_{\rm L}\approx R_{\rm out}$,
the wind expands adiabatically and radially, 
with a constant velocity $v_{\rm w}$, which we parametrize as a factor $f_{\rm v} \ge 1$ times 
 the escape velocity $\sqrt{2GM/r_{\rm L}}$,
 \be
 v_{\rm w} \approx 4.4 ~ 10^{9} f_{\rm v} \beta^{1/2} M_6^{1/3} x_{\star}^{-1/2} m_*^{1/6} \mbox{cm/s}.
 \ee
This is justifiable because $R_{\rm out}$ is close to the sonic radius,
after which the velocity varies only by a factor of a few.
We note that the two wind parameters, $f_{\rm v}$ and $f_{\rm out}$, should be determined and related by global conservation of energy and mass for the whole system.
However, given the level of simplification of our model, we do not attempt to calculate this relationship,
which depends on the details of how energy and mass are actually ridistributed between the disc and the wind.
For this reason, in the following, we will treat these parameters as independent.

The radiation is advected with the flow up to the trapping radius, where the optical depth is $\tau \simeq c/v_{\rm w}$.
Afterwards, the radiative transfer takes over, until photons reach the photospheric radius $r_{\rm ph}$. 
Since  $c/v_{\rm w} \approx 7$, the trapping radius 
and the photosphere are close enough that we may assume that the radiation keeps cooling along
the same adiabatic curve for the whole evolution. As we show in the following the temperatures in the flow are high enough 
to justify the use a constant Thomson opacity for a fully ionized gas ${\kappa}_{\rm s}=0.4$ g cm$^{-2}$.  

The radiation temperature at the base of the wind, $T_{\rm L}\equiv T(r_{L})$ can be derived from energy conservation in the wind. If all the internal energy at the base of the wind is converted into kinetic energy at large distances the conservation law can be written as
\be
4\pi r_{\rm L}^2 v_{\rm w} \left(\frac{4}{3} a T_{\rm L}^4\right) \simeq \frac{1}{2} \dot{M}_{\rm fb} f_{\rm out} v_{\rm w}^2,
\ee
from which we get
\be
T_{\rm L}(t_{\rm min}) \approx 3.5~ 10^{5}~ (f_{\rm out} f_{\rm v})^{1/4} \beta^{11/8} M_6^{-5/24}  x_{\star}^{-1} m_*^{17/24}~\mbox{K}.
\ee
 However, the wind is highly opaque and photons are released only
at a much larger radius,  the photospheric radius
\be
r_{\rm ph}(t_{\rm min}) \approx 1.4 ~10^{15}\frac{f_{\rm out}}{f_{\rm v}}  \beta^{5/2} M_6^{-5/6} x_{\star}^{-1} m_*^{11/6}~{\rm cm},
\ee
which is $\sim 100$ times $r_{\rm L}$. The corresponding photospheric temperature is
 $T_{\rm ph} =T_{\rm L} \left(r_{\rm ph}/r_{\rm L}\right)^{-2/3} \left(f_{\rm out}/f_{\rm v} \right)^{1/3}$, given by
\be
  T_{\rm ph}(t_{\rm min}) \approx \frac{1.6 ~ 10^{4}}{ \beta^{23/24}} {x_{\star}}^{1/3} \left(\frac{f_{\rm v}^{7}}{f_{\rm out}}\right)^{1/12}  \left(\frac{M_6^{41}}{m_*^{53}}\right)^{1/72} \mbox{K}.\\  
\ee
This gives an initial flash, most luminous in the optical band.
As the accretion rate decreases, the photosphere sinks inwards as
\be
r_{\rm ph} = r_{\rm ph}(t_{\rm min}) \left(\frac{t}{t_{\rm min}}\right)^{-5/3},
\ee
and the corresponding temperature {\em increases} in time as 
\be
 T_{\rm ph} =  T_{\rm ph}(t_{\rm min}) \left(\frac{t}{t_{\rm min}}\right)^{25/36}. 
\label{eq:tph}
\ee
If initially (at $t_{\rm min}$) the observed frequency $\nu_{\rm obs}$ lies much {\em below} the black body peak frequency, then the monochromatic lightcurve
decreases quite steeply as
\be
\nu_{\rm obs} L_\nu \propto r_{\rm ph}^2T_{\rm ph} \propto t^{-95/36}.
\label{eq:lw_rj}
\ee
The peak in the lightcurve is thus at $t=t_{\rm min}$,
\be
\begin{array}{ll}
\left .\nu_{\rm obs} L_{\rm \nu}\right |_{\rm t_{\rm min}} = \displaystyle \frac{8 \pi^2 \nu_{\rm obs}^3}{c^2} k_{\rm b} T_{\rm ph}(t_{\rm min}) r_{\rm ph}(t_{\rm min})^2 \approx \\
\nonumber                      3.7~ 10^{41} ~\nu_{\rm 14}^{3}\, \beta^{97/24}   \left(\displaystyle\frac{f_{\rm out}^{23}}{f_{\rm v}^{17}}\right)^{1/12} \left(\displaystyle\frac{m_*^{211}}{M_6^{79}}\right)^{1/72} x_{\star}^{-5/3} \mbox{erg/s},
\end{array}
\ee
\no
which, in the optical band, can be $10^{-2}-10^{-3}$ of the  Eddington luminosity.

When, instead, the observed frequency initially lies in the Wien part of the spectrum, then the
lightcurve has an initial exponential rise. The time of the peak is reached when $T_{\rm ph}$ (eq~\ref{eq:tph}) 
equals $T_{\rm obs}$ (Eq~\ref{eq:tobs}), 
\be
t_{\rm max} \approx 1.2 ~ 10^3~\left(\frac{\nu_{16}^{36}}{M_6^8}\right)^{1/25} \left(\frac{f_{\rm out}^{3}}{f_{\rm v}^{21}}\right)^{1/25} \left(\frac{m_*^3 x_{\star}^{51}}{\beta^{81}}\right)^{1/50} \mbox{d}
\ee
Afterwards, the lightcurve decreases as eq.~(\ref{eq:lw_rj}).
The peak luminosity in this case is
\be
\begin{array}{ll}
\left. \nu_{\rm obs} L_{\rm p,\nu}\right|_{\rm t_{\rm max}}  = 4 \pi r_{\rm ph}(t_{\rm max})^2 \sigma T_{\rm obs}^{4} = \\
= 1.5 ~ 10^{43} \nu_{16}^{-4/5}  m_*^{2/15}  \left(\displaystyle\frac{\beta}{ x_{\star}}\right)^{2/5} \left(\displaystyle\frac{f_{\rm out}^{8}}{f_{\rm v}^{24}}\right)^{1/5} M_6^{16/15} \mbox{erg/s}.\\
 \end{array} 
\ee
\no
If the observed frequency is in X-rays, at $0.2$ keV, the peak emission is $\approx 10^{41}$ erg/s at $t_{\rm max}\approx 9000$ days (for $f_{\rm out}=0.1$), which means that the wind contribution in the X-ray is much lower than the disc one (see Fig.~\ref{fig:discbolo}) for the whole duration of the wind $ t \le t_{\rm edd}$. Note that for the example above $t_{\rm max} \approx 9000~{\rm days} \gg t_{\rm edd}$ so the peak in the soft X-ray  lightcurve  is never reached, because the wind ceases before.

In reality, the fallbak rate may not be a pure power-law, but rather an exponentially increasing function, followed by a plateau which will then lead to the $t^{-5/3}$ decreasing slope. This has three main consequences. First, the duration of the super-Eddington phase (thus of the wind) may be shorter. Then, lower accretion rates are attained with respect to the power-law extrapolation at early times. Finally, there may be two spectral breaks in the lightcurve, since at early times, when $\dot{M}$ increases, the photospheric temperature {\em decreases}, while at late times, when $\dot{M}$ decreases, the opposite occurs. Therefore, it may be possible that the peak of the black body spectrum crosses twice the observed frequency.

\section{Total lightcurve}
\label{sec:lightcurve}

Having discussed the fundamental scalings of the two components of the
emission separately, we now turn our attention to the total lightcurve.
In particular, we discuss the consequences of relaxing two of our previous main
assumptions: (a) that the fallback rate
follows a simple power-law decline at all times and (b) that the gas
fraction ejected through a radiative wind $f_{\rm out}$ is constant,
independent on the fallback rate.
Also,  we will adopt a slim disc
configuration during the super-Eddington phase, where the
characteristic disc temperature is modified as:
\begin{equation}
\sigma_{\rm SB}T_{\rm d}^4 = \frac{3GM\dot{M}f}{8\pi R^3}\left[\frac{1}{2}+\left(\frac{1}{4}+\left(\frac{\dot{M}}{\eta\dot{M}_{\rm Edd}}\right)^2\left(\frac{R_{\rm s}}{R}\right)^2\right)\right]^{-1},
	\label{eq:t}
\end{equation}
\citep{strubbe09}, where $f=1-\sqrt{R_{\rm in}/{R}}$. The accretion rate in the disc is given by $\dot{M}=(1-f_{\rm out})\dot{M}_{\rm fb}$, where the fallback rate is computed either from the analytical expression (Eq. (\ref{eq:mdot})), or from the numerical results of \citet{LKP09}, depending on the models. From Eq. (\ref{eq:t}) we directly compute the disc effective temperature which allows us to compute the monochromatic luminosities as a function of time. 

\begin{figure*}
\begin{center}
\includegraphics[width=\columnwidth]{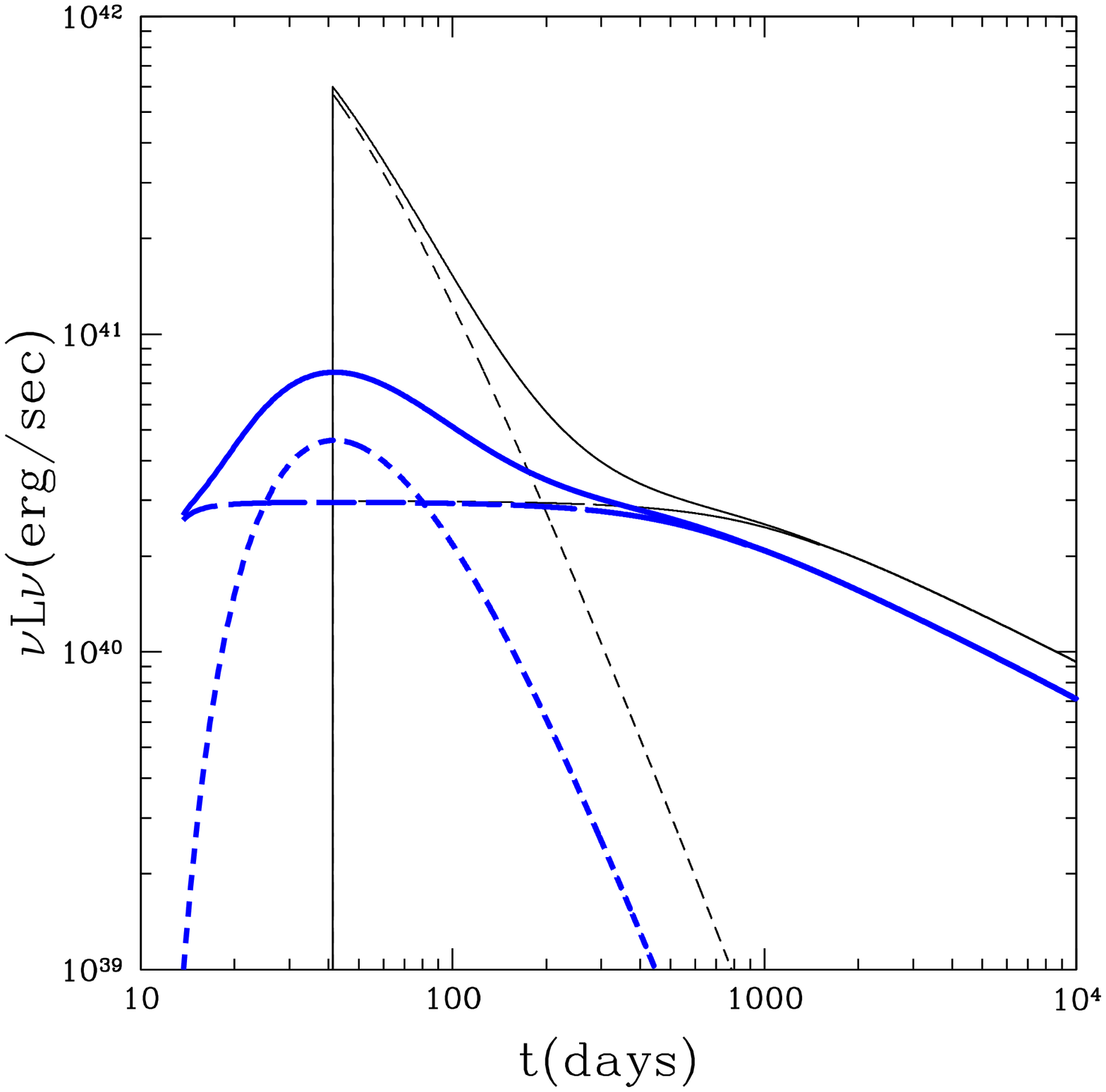}
\includegraphics[width=\columnwidth]{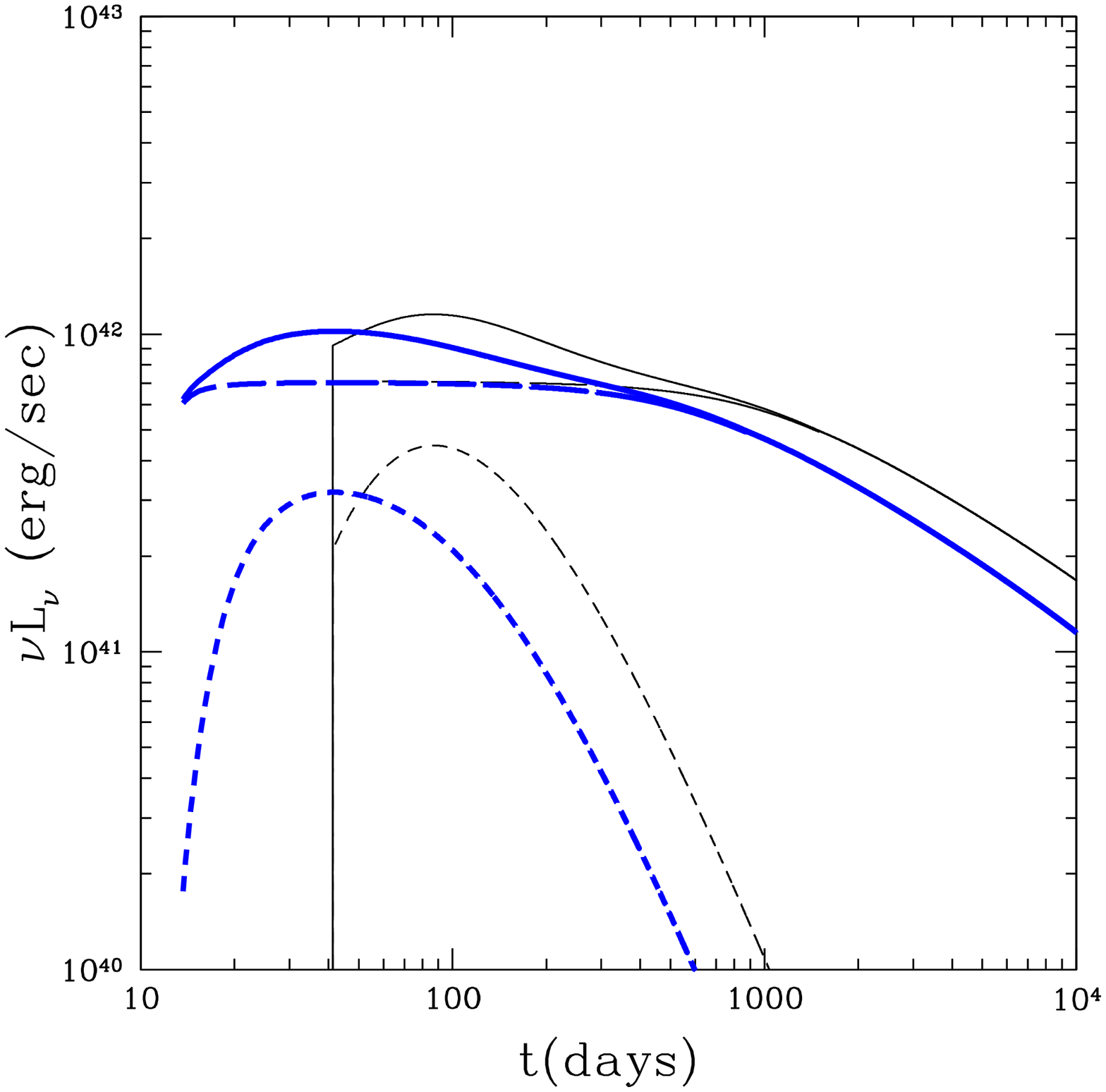}
\caption{G-band (left) and FUV (right) lightcurves of tidal disruption
  events, for a solar type star, $\beta=1$, a black hole mass $M=10^6M_{\odot}$, $f_{\rm out}=0.1$ and $f_{\rm v}=1$. 
  The black curves refer to a pure power law decline of the
  fallback rate, while the blue lines adopt the detailed modeling of
  the stellar structure of \citet{LKP09}, for $\gamma=1.4$. The solid lines are the
  total lightcurves, the short-dashed line indicates the wind
  component, and the long-dashed line the disc component. 
  In all cases, after $\approx 1$ year the disc dominates the
  emission. When a more detailed model for the fallback rate is used,
  which takes into account the internal structure of the disrupted
  star, the wind contribution is severlely suppressed and only
  provides a modest contribution in the optical. Lightcurves for the case of a $10^{5}M_{\odot}$ and of a
  $10^{7}M_{\odot}$ black holes are shown in Figs. \ref{fig:M5} and \ref{fig:M7}, respectively.
  }
\label{fig:M6}
\end{center}
\end{figure*}

\begin{figure*}
\begin{center}
\includegraphics[width=\columnwidth]{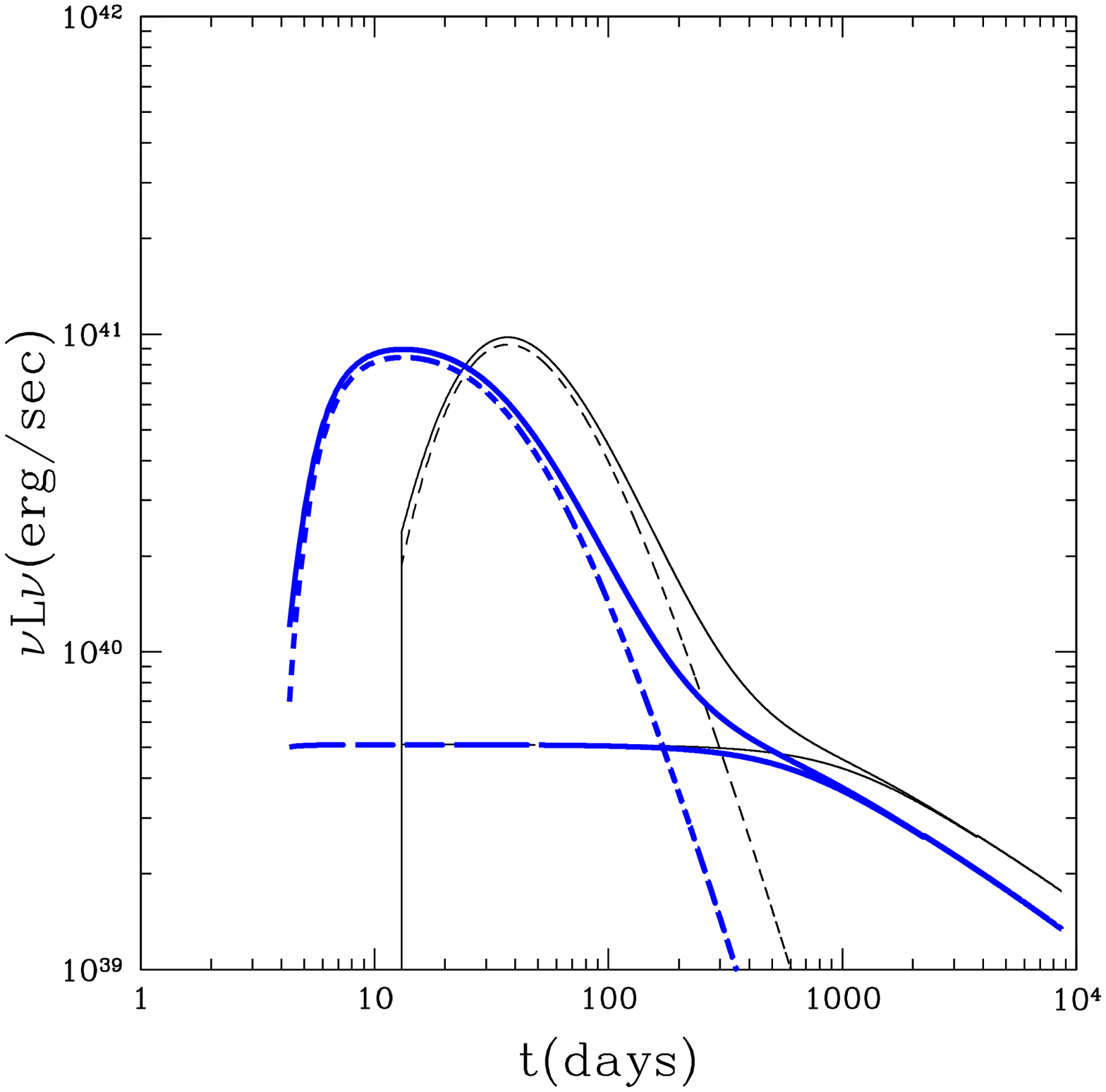}
\includegraphics[width=\columnwidth]{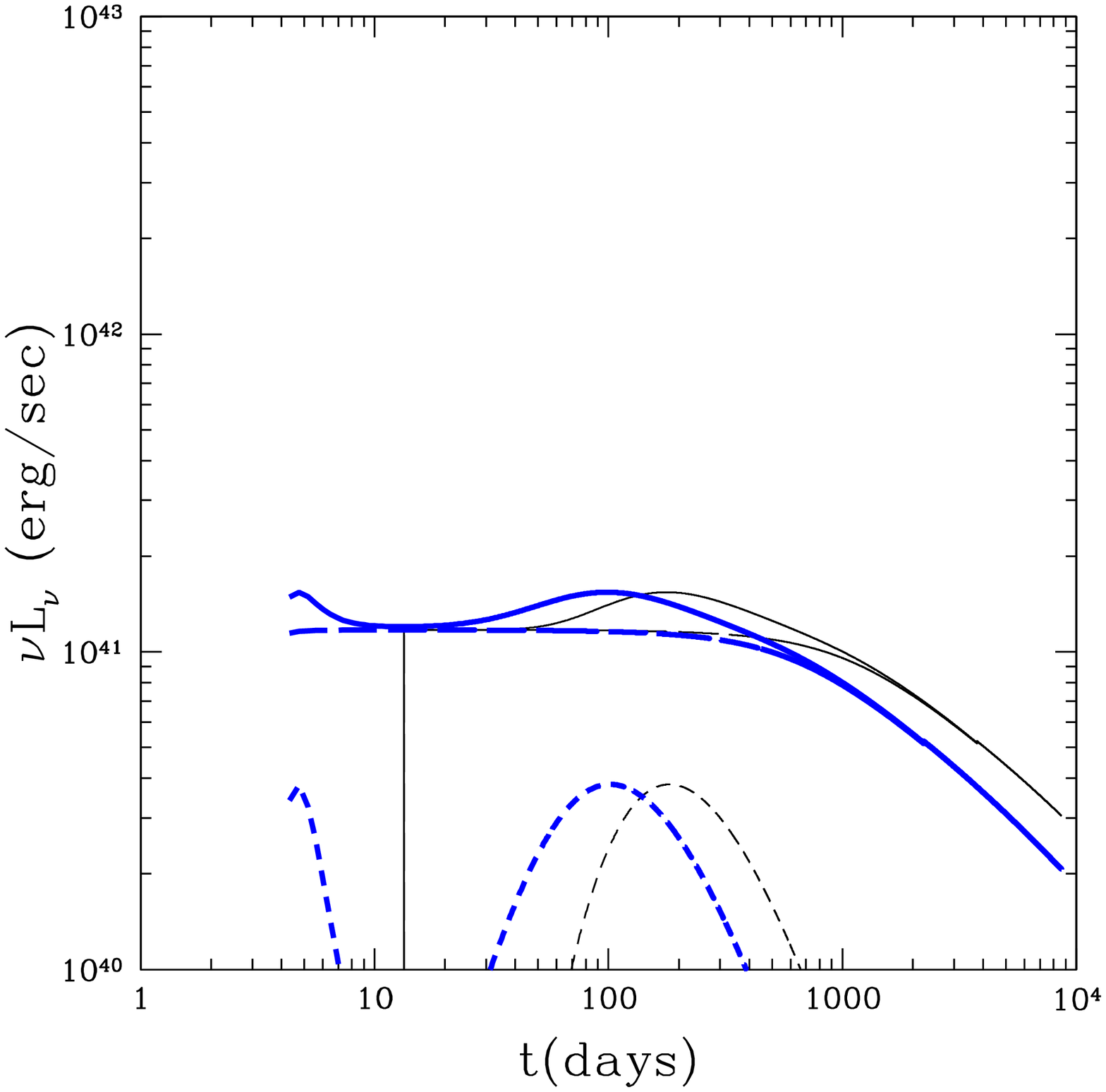}
\caption{Same as Fig. \ref{fig:M6}, but for $M=10^5M_{\odot}$.}
\label{fig:M5}
\end{center}
\end{figure*}

\subsection{Modeling the fallback rate: wind suppression}

As mentioned above, eqs. (\ref{eq:mdot}) and (\ref{eq:mdot2}), only
hold under the assumption that the specific energy distribution within
the disrupted star is constant. As shown by \citet{LKP09}, the actual
energy distribution depends on the stellar structure and only
approaches a constant value in the limit of a completely
incompressible star. If the star is compressible, the net effect is a
more gentle approach to the peak, with significant emission starting
at times slightly earlier than $t_{\rm min}$. The peak fallback rate
is consequently decreased with respect to the prediction of
Eq. (\ref{eq:mdot2}).  For example, if the star is modeled as a
polytropic sphere of index $\gamma=1.4$, the peak accretion rate 
for a black hole mass $M=10^6M_{\odot}$ is 
only $\dot{M}_{\rm p}\approx 40 \dot{M}_{\rm Edd}$, to be compared
with $\dot{M}_{\rm p}\gtrsim 100 \dot{M}_{\rm Edd}$ for the case of a
pure power-law decline. We thus expect that such a more detailed
modeling of the fallback rate would result in a suppression of the
wind contribution.

We computed the total lightcurves in several monochromatic bands
for the two extreme cases of an incompressible star (pure power law
decline in $\dot{M}_{\rm fb}$) and for the most compressible star
considered by \citet{LKP09}, with $\gamma = 1.4$. In particular, we
computed the fallback rate based on their equations (6)-(9),
including the homologous expansion by a factor $\approx 2.5$, as
observed in their numerical simulations.

Fig. \ref{fig:M6} shows the total lightcurves (solid lines), and the 
separate disc (long-dashed lines) and wind (short-dashed lines) components 
for $M=10^6M_{\odot}$. We also adopt  $f_{\rm out}=0.1$ and $f_{\rm v}=1$ for the wind (cf.
\citealt{strubbe09}). The left panel shows the g-band lightcurve, while the right panel shown the FUV one.
The black lines correspond to a pure power law decline
in the fallback rate, while the blue lines refer to the \citet{LKP09}
model. In all cases, after $\approx 1$ year the disc  dominates the emission in 
all bands, and the total lightcurve follows the $t^{-5/12}$ decline discussed in section
\ref{sec:scalings}. In the FUV band the disc contribution dominates at
all times. In the g-band, instead, the wind produces a luminous flare when the fallback rate is modeled
as a pure power-law. Note that our predicted lightcurve is slightly
different that the one presented in \citet{strubbe09} (their fig. 3,
middle panel). This is becasue they assume a wider distribution of
specific energies (by a factor 3), and because they have artificially
softened the sharp rise to the peak at $t=t_{\rm min}$ predicted in
this case (Strubbe, private communication). At late times, the wind
emission declines steeply with time as $t^{-95/36}$, as predicted by
the simple scalings of section \ref{sec:scalings}.

\begin{figure}
\begin{center}
\includegraphics[width=\columnwidth]{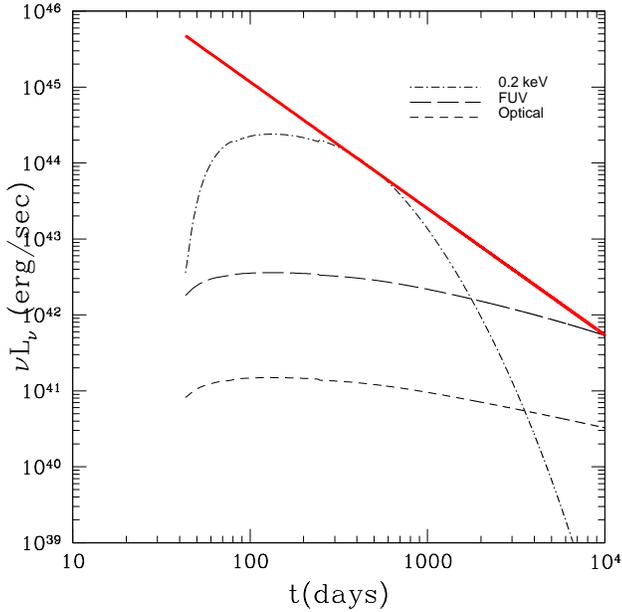}
\caption{G-band (short-dashed line), FUV (long-dashed line) and X-ray in the 0.2 keV band (dot-dashed line) lightcurves, for $f_{\rm out}=0.1$, $f_{\rm v}=1$, a black hole mass of $10^7M_{\odot}$, and our standard parameters $x_{\star}=m_*=\beta=1$ . The red line marks a $t^{-5/3}$ decline.}
\label{fig:M7}
\end{center}
\end{figure}

\begin{figure}
\begin{center}
\includegraphics[width=\columnwidth]{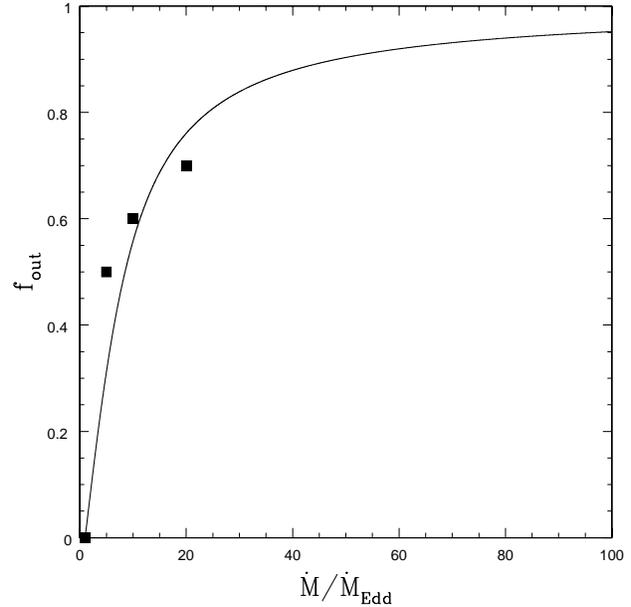}
\caption{The solid line shows the relation between $f_{\rm out}$ and $\dot{M}_{\rm fb}/\dot{M}_{\rm Edd}$ that we have used to model the super-Eddington accretion with a time-dependent fraction of ejected material (Eq. \ref{eq:dotan}). The squares indicate the results of \citet{dotan10}.}
\label{fig:dotan}
\end{center}
\end{figure}

As expected, the effect of a more gentle approach to the peak fallback
rate is to severely suppress the wind contribution, which, for a
polytropic star of index $\gamma=1.4$, provides only a modest
enhancement of the luminosity in the optical, while the disc emission
dominates at almost all times.

We have also computed the same lightcurves for a smaller black hole,
with a mass of $10^5 M_{\odot}$, and keeping all the other parameters
fixed. The results are shown in Fig. \ref{fig:M5}. For
such a smaller black hole, the fallback rate is more strongly
super-Eddington and thus the wind contribution is more prominent, even
when we adopt the \citet{LKP09} model.  In this case, the wind lightcurve
in the pure power-law model also shows an initial
 rise to the peak, before gradually declining as
$t^{-95/36}$. This is because $T_{\rm obs}$ initially lies in the
Wien part of the blackbody spectrum of the wind photosphere. As the
wind photosphere retreats and gets hotter, the observed frequency crosses the blackbody peak
and moves to the Rayleigh-Jeans side, thus producing the peak
in the monochromatic lightcurve (see section \ref{sec:scalings}).

Finally, we computed the lightcurves for a more massive
black hole, with $M=10^7 M_{\odot}$. In this case, the peak fallback
rate is only marginally super-Eddington. Even adopting a pure
power-law decline, we find that the wind provides a small contribution
in the optical (cf \citealt{strubbe09}). With the \citet{LKP09} model 
the peak fallback rate is
 $\dot{M}_{\rm p}\approx 1.3~\dot{M}_{\rm Edd}$. The wind is thus 
further suppressed and it produces a negligible
contribution at all wavelengths and at all times. The g-band,
FUV and X-ray lightcurves are
shown in Fig. \ref{fig:M7}. Even if here the disc dominates at all times,
the lightcurve is not proportional to $t^{-5/3}$ (the
red line in the figure) in any single band. Indeed, while in the
optical and in the FUV the lightcurve follows $t^{-5/12}$, in the X-rays it
initially grows, following the growth of $\dot{M}$, then it drops as
$t^{-5/3}$ for a short period of time, before eventually declining
exponentially.

We thus see that the shape of the wind lightcurve, which is strongly dependent on the evolution of the fallback rate at early times, is in turn related to the structure of the disrupted star.
More incompressible stars give rise to an initially larger fallback rate and are expected to produce a stronger wind contribution.

\subsection{Modeling super-Eddington accretion: wind enhancement}

\begin{figure*}
\begin{center}
\includegraphics[width=\columnwidth]{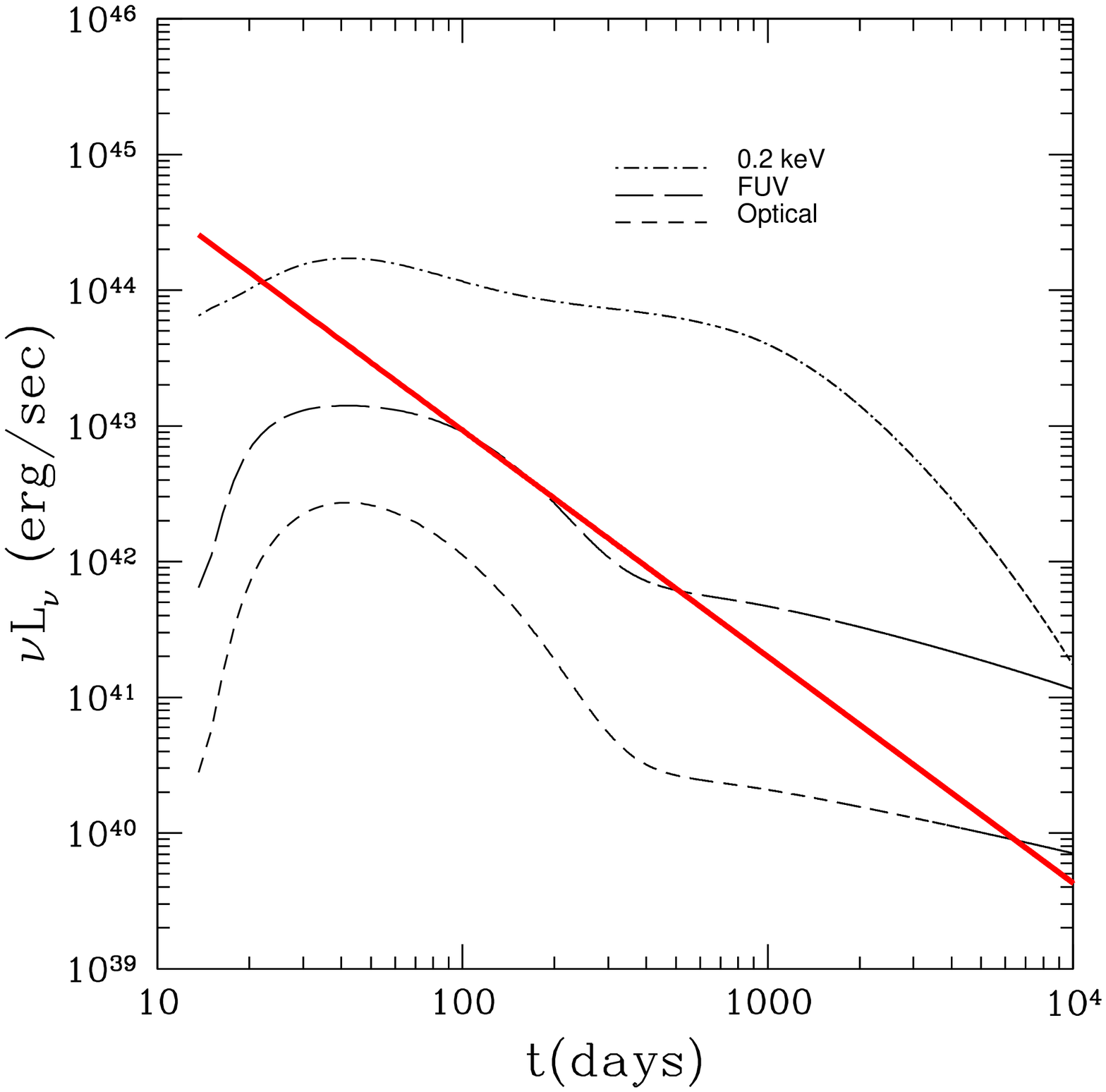}
\includegraphics[width=\columnwidth]{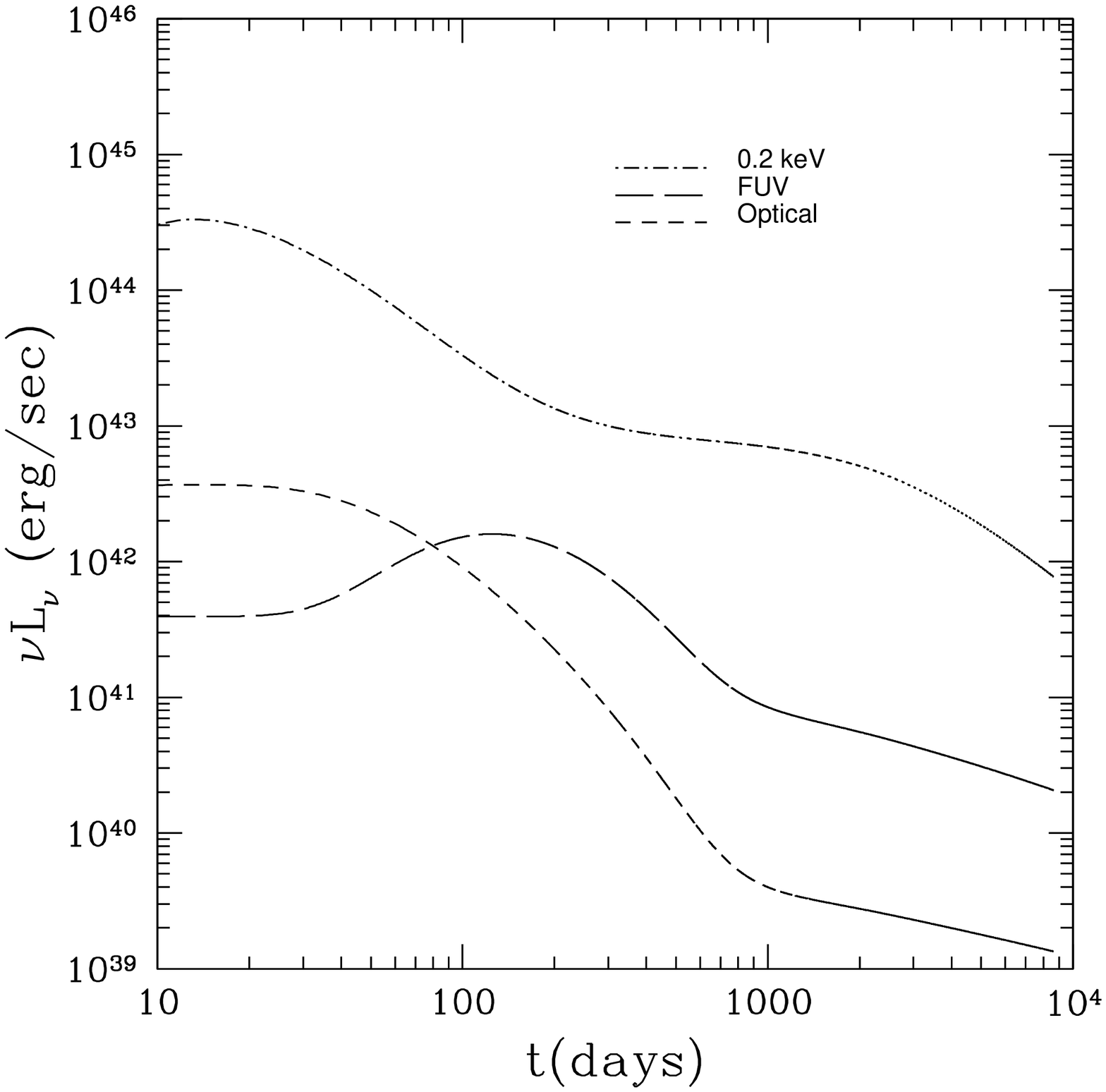}
\caption{Optical (g-band, short-dashed lines), FUV (in the GALEX FUV band, long-dashed lines), and X-ray (at 0.2 keV, dot-dashed lines) lightcurves obtained adopting a time-dependent model for $f_{\rm out}$ (eq. \ref{eq:dotan}), for $M=10^6M_{\odot}$ (left panel) and $M=10^5M_{\odot}$ (right panel). We also assume the \citet{LKP09} fallback rate for a $\gamma=1.4$ polytropic star. The other parameters are $x_{\star}=m_*=\beta=1$. The red line in the left panel shows, for comparison, a $t^{-5/3}$ decline. In this case, the wind contribution is strongly enhanced and now stands out both in the optical and in the FUV. On the contrary, the X-ray lightcurve is at all times dominated by the disc. The slight "bends" in the X-ray lightcurves are a consequence of the fact that, during the super-Eddington phase, only a (time-dependent) fraction of the fallback material makes it to the hole.}
\label{fig:fout}
\end{center}
\end{figure*}

The results presented in the previous section are in fact conservative estimates of the wind contribution. Indeed, we have assumed that only 10 percent of the infalling stellar debris are ejected in the radiatively driven wind, \emph{independently of the fallback rate}. In reality, the strength of the wind is expected to scale with the fallback rate, in such a way that very super-Eddington inflows should results in a much stronger wind. As mentioned above, it is not easy to quantify the fraction $f_{\rm out}$. We consider here the results of \citet{dotan10}, who have constructed self-consistent models for super-Eddington slim discs. These authors find that the fraction $f_{\rm out}$ is a growing function of the ratio $\dot{M}_{\rm fb}/\dot{M}_{\rm Edd}$, reaching values of the order of $f_{\rm out}\approx 0.7$ for $\dot{M}_{\rm fb}/\dot{M}_{\rm Edd}=20$. For larger values of the infall rate, they do not find any steady state solution, and it is thus not easy to extrapolate their results beyond that point. We note however, that, when the fallback rate is modeled based on the \citet{LKP09} results, the peak rate is not much larger than $20\, \dot{M}_{\rm Edd}$, and does not require an extrapolation much beyond their maximum value. 

\citet{dotan10} construct four different models of super-Eddington slim discs, with $\dot{M}_{\rm fb}/\dot{M}_{\rm Edd} = $ 1, 5 10 and 20, respectively. We have approximated these results using the following relation between $f_{\rm out}$ and $\dot{M}_{\rm fb}/\dot{M}_{\rm Edd}$:
\be
f_{\rm out} = \frac{2}{\pi}\arctan\left[\frac{1}{7.5}\left(\frac{\dot{M}_{\rm fb}}{\dot{M}_{\rm Edd}}-1\right)\right],
\label{eq:dotan}
\ee 
which has the required properties of approaching zero when $\dot{M}_{\rm fb}=\dot{M}_{\rm Edd}$ and of approaching unity when $\dot{M}_{\rm fb}\gg \dot{M}_{\rm Edd}$. A comparison between the relation predicted by Eq. (\ref{eq:dotan}) and the \citet{dotan10} results is shown in Fig. \ref{fig:dotan}. We have experimented with several different functional relations between $f_{\rm out}$ and $\dot{M}_{\rm fb}/\dot{M}_{\rm Edd}$ and found that, as long as the relation reasonably follows the \citet{dotan10} data, our resulting lightcurves are essentially unchanged.

Fig.  \ref{fig:fout} shows 
the monochromatic lightcurves obtained assuming the \citet{LKP09} fallback rate for a $\gamma=1.4$ polytropic star, for $M=10^6M_{\odot}$ (left panel) and $M=10^5M_{\odot}$ (right panel). Clearly, the wind contribution is strongly enhanced and now stands out at early times both in the optical and in the FUV for both choices of the black hole mass. On the other hand, in the X-rays, the disc contribution dominates at all times. After 1-3 years (depending on the black hole mass) the fallback rate drops to sub-Eddington values and the optical and FUV lightcurves approach the predicted $t^{-5/12}$ decline, while the X-rays decline exponentially. During the super-Eddington phase the shape of the lightcurve depends on the combination of various factors: the time-dependence of both $\dot{M}_{\rm fb}$ and of $f_{\rm out}$, the spectral changes occuring as the wind temperature increases, and the varying relative contribution of the disc and of the wind to the total emission. \emph{Coincidentally}, the FUV lightcurve for a $10^6 M_{\odot}$ black hole does initially show a $t^{-5/3}$ decline (shown with a red line in Fig. \ref{fig:dotan}) before flattening into $t^{-5/12}$.  

Finally, we considered the case of a more massive black hole, with $10^7M_{\odot}$. In this case the fallback rate is never much larger than the Eddington rate, and thus even when using Eq. (\ref{eq:dotan}) for $f_{\rm out}$ there is no strong wind, and the emission is at all times completely dominated by the disc.

\section{Conclusions}
\label{sec:discussion}

The `signature' of a tidal disruption event is traditionally considered to be a characteristic evolution of the {\em fallback rate}, which should follow approximately a $t^{-5/3}$ decline.
Indeed, in most cases candidate tidal disruption events are identified by observing a $t^{-5/3}$ decay in the  {\em luminosity} over the course of a few months at various wavelengths, from the UV \citep{gezari08,gezari09} to the X-rays \citep{esquej08,cappelluti09}. In this paper,  we have critically reviewed the assumptions that
 i) the fallback/accretion rate should follow at all times a $t^{-5/3}$ decay and ii) that its behaviour directly translates into the observed lightcurve decay.   
 
 As to the former, the fallback rate behaviour is expected to depend on the star internal structure  \citep{LKP09}.
 Generally, it displays initially a gentle rise to the peak and approaches a $t^{-5/3}$ decay only several months after the event. In this early stages, the rate is super-Eddington for a black hole of $M < 10^{7} M_{\odot}$, and the fraction of mass that accretes onto the hole or is lost in a radiatively driven wind is also time dependent \citep{dotan10}.  Under these conditions, the optical and UV lightcurves may display an intial bump given by the wind emission, which decay rapidly as $t^{-95/36}$. For more massive black holes, the wind is suppressed, since all or nearly all the fallback mass can be accreted, and the lightcurve at all frequencies is dominated by the disc emission. The detailed evolution and the strength of the wind emission is strongly dependent on the evolution of fallback rate at early times, and is thus very sensitive to the internal structure of the disrupted star \citep{LKP09}
 
Likewise, at higher frequencies and at later times, the wind emission is negligible for all black hole masses.  Also in this cases, while the bolometric curve may decline as $t^{-5/3}$, the monochromatic lightcurves in general do not. On a timescale of months to years, the optical and UV lightcurves are expected to decrease much slower, as $t^{-5/12}$. Only in X-rays does the lightcurve initially follow more closely the bolometric luminosity, hence showing an approximately $t^{-5/3}$ decline, before steepening into an exponential drop after a few years. The predicted X-ray luminosities in the first year after the disruption are comparable to those of bright quasars. Detectability is thus not an issue in the local universe, and a standard snapshot survey with currently flying detectors would give enough accuracy to constrain the slope with two epochs of observation. Coincidentally, for some combinations of the model parameters, the FUV lightcurve might exhibit a decline seemingly close to $t^{-5/3}$, although clearly this does not reflect the time evolution of the fallback rate.
 
Typically, observed tidal disruption events candidates have UV/optical luminosities in excess of $10^{43}$ ergs/sec and lightcurves which decrease more steeply that $t^{-5/12}$. This indicates that even in the FUV the wind contribution might dominate at early times, consistent with our models which adopt a time dependent ejected mass fraction (see, for example, Fig. 7, left panel).   We conclude that a more sophisticated --- ideally multi-wavelength --- modelling of the data than done so far is required to pin down a tidal disruption event and derive the system parameters. The X-ray band seems best suited to identify a tidal disruption event through the fallback decay rate, while early optical and UV observations can catch at early times the wind emission, which would also teach us about the ill-known super-Eddington accretion physics and possibly even about the structure of the disrupted star.

\section*{Acknowledgements}

We would like to thank Linda Strubbe and Eliot Quataert for stimulating discussions about their work.  GL acknowledges the hospitality of the Racah Institute of Physics in Jerusalem, where a large part of this work was done.

\bibliography{lodato}

\label{lastpage}
\end{document}